# Recent developments in the techniques of controlling and measuring suction in unsaturated soils

Pierre Delage[1] & Enrique Romero[2] and Alessandro Tarantino[3]

[1] Ecole des Ponts, CERMES (Université Paris-Est, UR Navier), France
[2] Universitat Politécnica de Catalunya, Barcelona, Spain
[3] Universita degli Studi di Trento, Italy

ABSTRACT: The difficulty of measuring and controlling suction in unsaturated soils is one of the reasons why the development of the mechanics of unsaturated soils has not been as advanced as that of saturated soils. However, significant developments have been carried out in the last decade in this regard. In this paper, a review of some developments carried out in the techniques of controlling suction by using the axis translation, the osmotic method and the vapour control technique is presented. The paper also deals with some recent developments in the direct measurement of suction by using high capacity tensiometers and in the measurement of high suction by using high range psychrometers. The recent progresses made in these techniques have been significant and will certainly help further experimental investigation of the hydromechanical behaviour of unsaturated soils.

## 1 INTRODUCTION

The coupled effects of changes in suction and stress on the response of unsaturated soils is a fundamental aspect to consider when dealing with unsaturated soils. The difficulty of measuring and controlling suction is one of the reasons why the development of the mechanics of unsaturated soils has not been as advanced as that of saturated soils in which water pressure is positive. In relation with the significant increase in research efforts carried out during the last two decades in the mechanics of unsaturated soils, various techniques of measuring and controlling suction have been adopted and/or further developed. These techniques have been described in detail in various papers (including Ridley and Wray 1996, Agus and Schanz 2005, Rahardjo and Leong 2006).

Recently, significant advances have been performed in the field of controlling and measuring suction. This paper deals with some recent achievements gained in the use of the three techniques of controlling suction, i.e. the axis-translation technique, the osmotic technique and the vapour control technique. Two techniques of measuring suction are also considered, i.e. high capacity tensiometers and high range psychrometers.

## 2 TECHNIQUES OF CONTROLLING SUCTION

### 2.1 Axis translation technique

#### 2.1.1 Introduction

The axis translation technique is the most commonly used technique of controlling suction. Early developments of this technique started with the pressure plate outflow technique (Richards 1941, Gardner 1956). The axis translation technique is associated with the matrix suction component, in which water potential is controlled by means of liquid phase transfer through a saturated interface –usually a saturated high air-entry value (HAEV) ceramic disk or a cellulose acetate membrane– which is permeable to dissolved salts. The procedure involves the translation of the reference pore air pressure, through an artificial increase of the atmospheric pressure in which the soil is immersed. Consequently, the negative pore water pressure increases by an equal amount if incompressibility of soil particles and water is assumed –i.e., if the curvature of the menisci is not greatly affected–. The translation of the pore water pressure into the positive range allows its measurement (Hilf 1956), and consequently, its control if water pressure is regulated through a saturated interface in contact with the sample. This technique has been experimentally evaluated with soils having a continuous air phase and a degree of saturation varying between 0.76 and 0.95 by Fredlund and Morgenstern (1977) and by Tarantino et al. (2000) for degrees of saturation between 0.56 and 0.77.

The axis translation technique has been criticised concerning the following aspects: i) it is not representative of field conditions where air pressure is under atmospheric conditions; ii) there are some doubts in how the air pressurisation process affects the water pressure when water is held by adsorption mechanisms; and finally iii) its application at nearly saturated states in the absence of a continuous gaseous phase is not straightforward. Nevertheless, the axis translation technique has proved to provide reasonable results and a good continuity between vapour equilibrium results at elevated suctions and nearly saturated states. An example can be found in Figure 1, in which the overall picture of water retention results under constant volume conditions obtained by combining different techniques (high-range transistor psychrometers and vapour control technique) jointly with axis translation, shows an adequate overlapping.

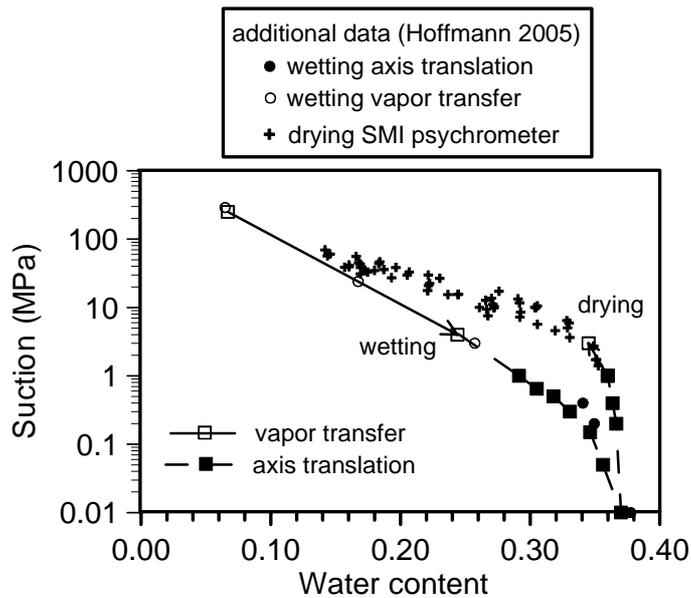

Figure 1. Water retention curves obtained by combining axis translation with other techniques (Hoffmann et al. 2005).

The major experimental difficulties concerning the application of the axis translation are associated with: i) the accumulation of diffused air beneath the HAEV ceramic disk, ii) the control of the relative humidity of the air chamber to minimise evaporation or condensation effects on the sample, iii) the application of the air pressurisation process at elevated degrees of saturation, and iv) the estimation of the equalisation time.

### 2.1.2 *Air diffusion*

Air diffusion through the saturated porous network of the interface can induce the progressive loss of continuity between the pore water and the water in the control system. In addition, the accumulation of air can lead to water volume change errors in drained tests and to pore-water pressure measurement errors in undrained tests. Consequently, an auxiliary device is required to flush periodically air bubbles accumulated below the HAEV ceramic. The following expression describes the rate of accumulation of dissolved air beneath the ceramic disk, which is based on the gradient of air concentration being the driving mechanism (Fredlund and Rahardjo 1993, Romero 1999):

$$\frac{dV_d}{dt} = \frac{n\,A\,D\,h\,(u_a - u_w)}{(u_w + u_{atm})\,t_c} \qquad (1)$$

where $n$, $A$ and $t_c$, represent the porosity, the cross-sectional area and the disk thickness, respectively. $h$, is the volumetric coefficient of solubility of dissolved air in water ($h=0.018$ at 22°C). $D$, is the diffusion coefficient through the saturated interface. $u_{atm}$, represents the absolute atmospheric pressure; $u_a$ and $u_w$ refer to air and water gauge pressures respectively. The quantification of air diffusion has been recently carried out by Romero (2001a), De Gennaro et al. (2002), Airò Farulla and Ferrari (2005) and Padilla et al. (2006). Lawrence et al. (2005) presented a pressure pulse technique for measuring the diffused air volume by using pressure/volume controllers.

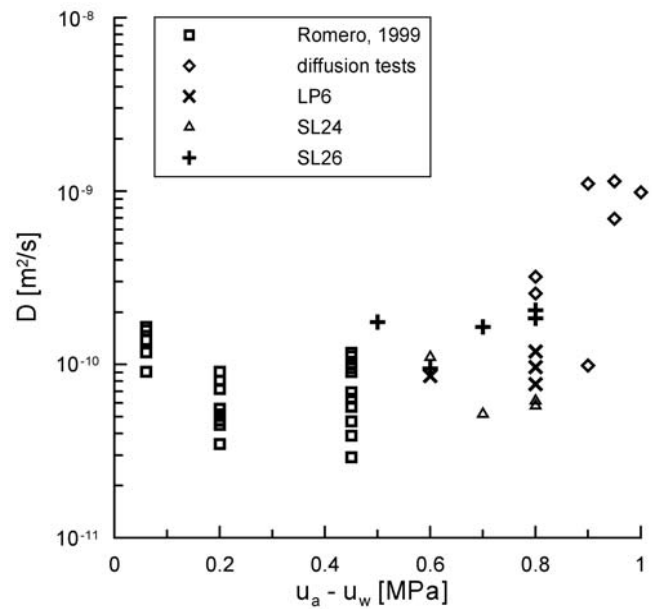

Figure 2. Diffusion coefficients for air through saturated ceramic disks as a function of the applied matrix suction (Airò Farulla and Ferrari 2005).

Figure 2 presents values of the coefficient of diffusion of air through a saturated ceramic disk with an air-entry value higher than 1 MPa as a function of the applied matrix suction. Typical values are included between $3\times10^{-11}$ and $2\times10^{-10}$ m$^2$/s (for suc-tions < 0.7 MPa) with lower values than that of air diffusion in water (around $2.2\times10^{-9}$ m$^2$/s at 20°C). Factors such as the tortuosity of the paths and a possible breakdown of Henry's law in curved air-water interfaces can be associated with this reduction (Barden and Sides 1967). The figure shows how this coefficient tends to increase as suction increases over 0.7 MPa and how it gets closer to the air-entry value of the ceramic (the value at which the gas convection transport is initiated). As deduced from Equation (1), increasing the water pressure is an efficient way to reduce air diffusion rates for a given geometry of the interface element and for a specified matrix suction. The conventional technique of the pressure plate apparatus, in which the pressure of water is maintained under atmospheric conditions, is the less efficient configuration to control the diffusion of air.

### 2.1.3 Evaporation and condensation effects

Vapour transfers between the soil and the surrounding air can be controlled by maintaining an adequate relative humidity in the air chamber (around 95%). Evaporative fluxes are originated by the difference in vapour pressure between the soil surface and the air chamber. Volumetric evaporative fluxes can be detected in the water volume change device as a non-stop inflow into the soil under steady-state conditions. Condensation of vapour in the internal walls of the pressure chamber due to temperature variations has also been reported by Oliveira and Marinho (2006).

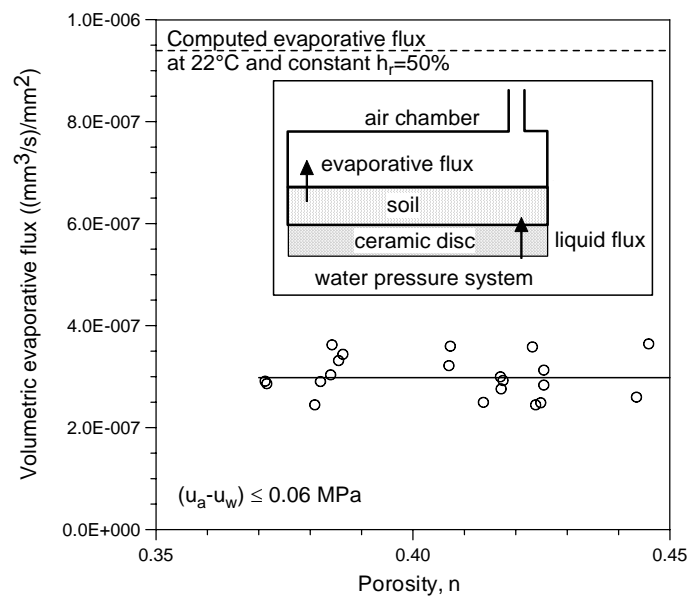

Figure 3. Measured volumetric evaporative fluxes (Romero 1999).

Measured volumetric evaporative fluxes at different porosities are presented in Figure 3 for a compacted clay specimen placed inside an air chamber at an initial relative humidity of 0.50 (Romero 1999). As shown in the figure, two different water fluxes are involved in the process: a) an evaporative flux that dries the clay surface, and b) a liquid flux through the ceramic disk that regulates the imposed matrix suction. A series of 1-D numerical analysis was carried out by Romero (1999) to simulate evaporative fluxes and matrix suction changes during a wetting path. A maximum volumetric evaporative flux of $9.4\times10^{-7}$ (mm$^3$/s)/mm$^2$ was computed when an initial relative humidity of 0.5 was imposed in the air chamber. With measured volumetric evaporative fluxes lower than this value, no important consequences are expected and relatively uniform matrix suction distribution is expected throughout the sample height (Romero 1999, 2001a,b).

### 2.1.4 Air pressurisation at elevated degrees of saturation

The application of air pressure at elevated degrees of saturation (involving occluded air bubbles) can induce irreversible arrangements in the soil skeleton due to pore fluid compression and to the fact that air pressure acts as a total stress when the continuity of air is not ensured. Bocking and Fredlund (1980) studied the effect of occluded air when using the axis translation technique. As a consequence, if nearly saturated states are expected to be reached during the hydraulic paths, it is preferable to increase the air pressure when the continuity of air is ensured (degrees of saturation < 0.85) and then to maintain the continuous air phase at constant pressure. After this initial stage it is possible to attain nearly saturated states, since the air pathways have already been created. This can be observed in Figure 1, in which the drying path followed a wetting path that attained very low matrix suctions. Nevertheless, if air pressure is required to be increased at high saturation values, it is preferable to change it at very slow rates to allow the system to create air pathways and to diffuse air through the liquid (Di Mariano 2000, Romero 2001a).

### 2.1.5 Time to reach suction equalisation

An important difficulty faced when using the axis translation technique is the estimation of the required time to reach suction equalisation. Water volume measurements are usually affected by the relative humidity of the air chamber and the diffusion of air. Although these phenomena can be minimised as previously suggested, the estimation of the equalisation time in oedometer and triaxial cells has been conventionally determined based on overall soil volume change measurements that are independently determined. Oliveira and Marinho (2006) studied the equilibration time in the pressure plate and recommended around three days for increments from 50 kPa to 100 kPa for gneissic soils.

From the analytical solution proposed by Kunze and Kirkham (1962), that considers the ceramic disk impedance and the soil permeability to determine the time evolution of the water volume change in a soil with a rigid matrix, it is possible to estimate an equalisation time $t_{95}$ for which 95% of the water outflow or inflow has occurred (note that for simplicity only one term of the Fourier series has been kept):

$$t_{95} \approx -\frac{L^2}{\alpha_1^2 D} \ln\left[\frac{\alpha_1^2}{40}(a + \csc^2 \alpha_1)\right];$$
$$a\alpha_1 = \cot \alpha_1 \text{ with } 0 < \alpha_1 \leq \frac{\pi}{2};$$
$$D = \frac{k_w}{n\gamma_w}\frac{\delta s}{\delta S_r}$$

(2)

where $L$ is the soil height, $D$ the capillary diffusivity that is assumed constant and dependent on the water permeability $k_w$ and on the soil water capacity, $\delta s/\delta S_r$ (being $s$ the matrix suction, $S_r$ the degree of saturation respectively), $n$ the porosity and $\gamma_w$ the unit weight of water; $a$ the ratio of impedance of the ceramic disk with respect to the impedance of the soil $a = k_w t_c/(L k_d)$ (being $t_c$ the ceramic disk thickness and $k_d$ its water permeability respectively) and $\alpha_1$ the solution of the equation in the indicated range. For low disk impedance, $a \approx 0$ and $\alpha_1 \approx \pi/2$, the minimum equalisation time can be approximately estimated as:

$$t_{95} \approx 1.129 \frac{L^2}{D}$$ (3)

For a clayey soil with $L = 20$ mm, $n = 0.48$, $k_w = 5 \times 10^{-12}$ m/s and $\delta s/\delta S_r \approx 2.8$ MPa in the suction range $0.1$ MPa $< s < 0.5$ MPa and with disk properties characterised by $t_c = 7$ mm and $k_d = 10^{-10}$ m/s, then $a \approx 0.018$, $\alpha_1 \approx 1.543$, $D \approx 3.0 \times 10^{-9}$ m$^2$/s and $t_{95} \approx 2615$ min. If no ceramic disk impedance is considered, then $t_{95} \approx 2500$ min. It is important to remark that this estimation is based on the hypothesis of a constant soil volume, which is not exactly the case with a clayey soil. Nevertheless, it gives an approximate estimation of the minimum time required to reach suction equalisation.

As a conclusion, provided its specific problems are adequately considered, the axis translation method has proven to be an efficient and reliable technique of controlling suction. It remains widely used to determine the water retention and transfer properties and the mechanical behaviour features of unsaturated soils, following the first adaptation to triaxial testing by Bishop and Daniel (1961).

## 2.2 Osmotic technique

### 2.2.1 Introduction

In the osmotic technique (see Delage and Cui 2008a) the sample is placed in contact with a semi-permeable membrane (permeable to water) while an aqueous solution containing large sized soluble polyethylene glycol molecules (PEG) is circulated behind the membrane. The PEG molecules cannot go through the membrane, resulting in an osmotic suction applied to the sample through the membrane. Being the membrane permeable to the salts dissolved in the water, the osmotic technique controls the matrix suction, like the axis translation technique. The value of the imposed suction depends on the concentration of the solution, the higher the concentration, the higher the suction. The suction/concentration relation will be discussed later in some details.

Semi-permeable membranes are characterised by there molecular Weight Cut Off (MWCO) that is linked to the size of the PEG molecules that they can retain (MWCO 12 000-14 000 membranes are used with PEG 20 000, MWCO 6 000 with PEG 3 500, MWCO 4 000 with PEG 2 000 and MWCO 1 500 with PEG 1 000). Note that the smaller the MWCO, the higher the membrane permeability. Semi-permeable membranes are most often made up of cellulose acetate, but interesting results using more resistant polyether sulfonated membranes have been published by Slatter et al. (2000) and Monroy et al. (2007). Semi-permeable membranes are obviously thinner than the ceramic disks used in the axis translation technique, but, as shown in Delage and Cui (2008a), they have comparable impedances $I$ ($I = e/k$, being $e$ and $k$ the thickness and water permeability respectively). On a Spectrapor 12 000-14 000 membrane, Suraj de Silva (1987) obtained $e = 50$ μm and $k = 10^{-12}$ m/s, giving $I = 5 \times 10^7$ s, compared to an impedance value of $7.5 \times 10^7$ s given by Fredlund and Rahardjo (1993) for a 6 mm thick 1 500 kPa air entry value ceramic disk.

The osmotic technique was used to control the osmotic pressure of culture solutions in biology by Lagerwerff et al. (1961) and the water matrix potential in soil science by Painter (1966), Zur (1966) and Waldron and Manbeian (1970). Peck and Rabbidge (1969) designed an osmotic tensiometer for measuring the osmotic suction applied as a function of the PEG concentration. The first application to geotechnical engineering was by Kassiff and Ben Shalom (1971) with subsequent work carried out on a hollow cylinder triaxial apparatus by Komornik et al. (1980) and on a standard triaxial apparatus by Delage et al. (1987). The Kassiff and Ben Shalom's device (Figure 4), was further improved by Delage et al. (1992) with the introduction of a closed circuit comprising a 1 litre bottle in which the solution was cir-culated by a peristaltic pump (being the bottle placed in a temperature controlled bath to allow water exchange measurements by using a capillary tube). The close circuit was adopted by Dineen and Burland (1995) with the bottle being permanently weighted by an electronic balance to monitor the water exchanges. Tarantino and Mongiovi (2000) and Monroy et al. (2007) also used this device.

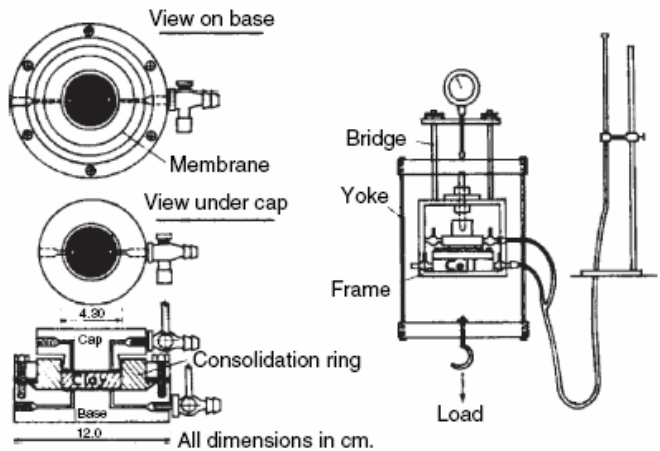

Figure 4. The osmotic oedometer of Kassif and Ben Shalom (1971).

Figure 5 shows a comparison carried out on a kaolinite slurry submitted to changes in suction over a wide range by using various suction control techniques, with a reasonable agreement observed between the various techniques. Ng et al (2007) drew similar conclusions based on results from shear testing.

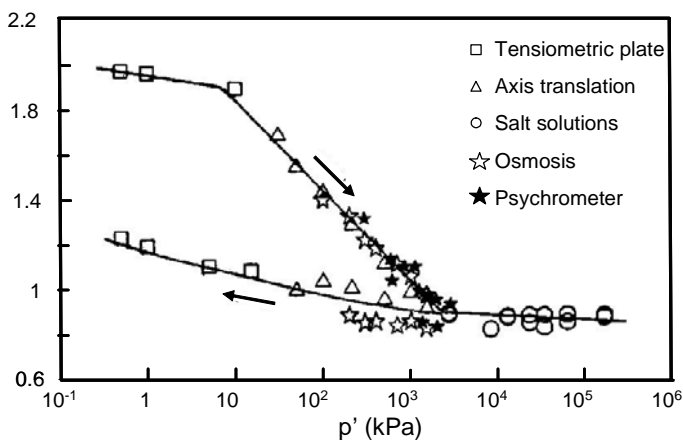

Figure 5. Comparison of the osmotic technique with various other suction control techniques (after Fleureau et al. 1993).

### 2.2.2 Advantages and drawbacks

Compared to the axis translation technique, the osmotic technique presents the advantage of exactly reproducing the real conditions of water suction, with no artificial air pressure applied to the sample. This advantage is believed to be significant in the range of high degrees of saturation when air continuity is no longer ensured with the apparition of occluded air bubbles and possible artefacts created by the air pressure application (see discussion above).

A technological advantage of the osmotic technique is that there is no need to apply any air pressure (resulting in no air diffusion problems). High level of suctions can easily be applied by using high concentration PEG solutions. It has been showed that the higher limit of the technique could be extended up to around 10 MPa (Delage et al. 1998) and an osmotically suction controlled oedometer compression test at a suction of 8.5 MPa has been presented by Cuisinier and Masrouri (2005a). This extension to high suction is obviously easier than when using the axis translation technique (Escario and Juca 1989).

In the triaxial apparatus, the application of high suctions is facilitated by the fact that there is no need to impose high values of confining stress to maintain constant the net total mean stress $\sigma - u_a$ at the elevated air pressures needed to impose high suctions. The highest suctions applied in triaxial testing (1500 kPa) were by using the osmotic technique (Cui and Delage 1996). In clays, with air entry values frequently higher than 1 MPa, this advantage is significant to ensure significant sample desaturation. In the oedometer, the application of the osmotic technique is easy since no air-tight device is necessary to apply the air-pressure on the sample, resulting in less friction effects between the piston and the ring. The adaptation of the osmotic technique to the oedometer only consists in replacing a porous stone (most often the bottom one) by a semi-permeable membrane clamped between the oedometer base and the ring. In the triaxial apparatus, the adaptation is less straightforward, as compared to that of the axis-translation technique, more often used.

The main drawback of the osmotic technique is the sensitivity to bacteria attacks of the cellulose acetate membranes that have been most commonly used up to now. When a semi-permeable membrane fails, the PEG solution can infiltrate the sample and suction is no longer controlled. The problem seems to be more serious when applying high suctions along wetting paths (suction decrease), as observed by Marcial (2003). In this regard, note that a concern recently evidenced by Delage and Cui (2008b) is related to the possible presence of PEG molecules of dimensions smaller than that defined by the molecular weight given by the manufacturer. The presence of these small molecules was demonstrated by developing a novel filtration system applied to PEG 6 000, filtrated by using a MWCO 3 500 cellulose acetate membrane.

When using cellulose acetate membranes, this effect can be corrected by adding few drops of penicillin in the solution. In such conditions (Kassif and

Ben Shalom 1971), the life duration of the membrane appears to be longer than 10 days. More recently, Slatter et al (2000) suggested the alternative use of polyether sulfonated semi-permeable membranes. By using these membranes, Monroy et al. (2007) carried out tests as long as 146 days. This option seems to be an excellent way to enhance the reliability of the osmotic method.

### 2.2.3 *Calibration of the method*

Initially, the calibration curves giving the total suction as a function of the solution concentration of various PEGs were investigated by measuring the relative humidity above solutions of PEG by using psychrometers (Lagerwerff et al. 1961, Zur 1966). The data from various authors gathered by Williams and Shaykewich (1969) indeed showed no significant difference between points obtained with PEG 6 000 and with PEG 20 000, the calibration curve being independent on the molecular mass of the PEG used. Based on this calibration, reasonable comparison with the axis translation techniques have been obtained by Zur (1966) and Waldron and Manbeian (1970) on various soils. This is confirmed by the data of Figure 5.

Further calibrations were carried out by Dineen and Burland (1995) who used the high range tensiometer (up to 1500 kPa) developed by Ridley and Burland (1993). They made direct suction measurements on a sample kept under a suction controlled by the osmotic technique in a oedometer and they also measured directly the suction by placing the probe in contact, through a kaolinite thin layer, with the semi-permeable membrane behind which the solution was circulated. The same approach was adopted by Tarantino and Mongiovi (2000) and, more recently, by Monroy et al. (2007).

The effect of the pair membrane/PEG used on the calibration has been observed by these authors on various membranes and PEGs, as seen in Figure 6.

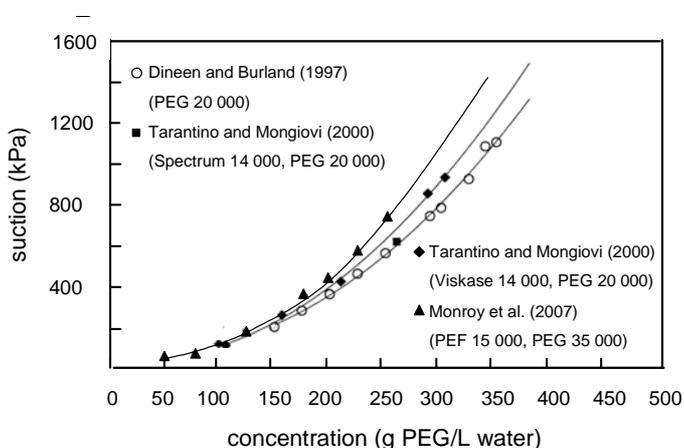

Figure 6. Dependency of the calibration curve of the osmotic technique with respect to the membrane and PEG used (after Delage and Cui 2008a).

In accordance with Slatter et al. (2000), Monroy et al. (2007) observed that, for a given concentration, the highest suctions were obtained by using the polyether sulfonated membrane with PEG 35 000 with suction values close to that of Williams and Shaykewich (1969). Monroy et al. (2007) also observed some difference when comparing calibration points along a wetting path (suction decrease) compared to that along a drying path (suction increase) with smaller suction obtained during the subsequent drying path. Actually, a similar membrane effect had also been observed from the data of Waldron and Manbeian (1970) who developed a null type osmometer in which the osmotic pressure was compensated by an air pressure applied to the solution for suctions included between 16 and 2480 kPa.

As a conclusion, it seems that the use of more resistant membranes together with the completion of specific calibrations based on the couple membrane/PEG used will give good reliability to the osmotic technique. The advantages of the technique should probably help for better experimental investigation and understanding of the transition zone at high degrees of saturation ($S_r < 0.85$), where the air continuity no longer stands and where samples get closer to saturation. The technique seems also particularly suitable to study the behaviour of unsaturated plastic soils with AEV higher than 1 MPa.

### 2.3 *Vapour control technique*

Vapour equilibrium technique is implemented by controlling the relative humidity of a closed system. Soil water potential is controlled by means of the migration of water molecules through the vapour phase from a reference system of known potential to the soil pores, until equilibrium is achieved. The thermodynamic relation between total suction of soil moisture and the relative humidity of the reference system is given by the psychrometric law (Fredlund and Rahardjo 1993). The relative humidity of the reference system can be controlled by varying the chemical potential of different types of aqueous solutions (Delage et al. 1998, Tang and Cui 2005).

Oedometer cells installed inside a chamber with relative humidity control were used by Esteban (1990), Bernier et al. (1997) and Villar (1999) and Cuisinier and Masrouri (2005b). The main drawback of this experimental setup is that the time to reach moisture equalisation is extremely long due to the fact that vapour transfer depends on diffusion (several weeks are required for each suction step in the case of high-density clays as observed in Figure 7). In order to speed up the process, vapour transfer – through the sample or along the boundaries of the sample– can be forced by a convection circuit driven by an air pump (Yahia-Aissa 1999, Blatz and Graham 2000, Pintado 2002, Lloret et al. 2003, Oldecop and Alonso 2004, Dueck 2004, Alonso et al. 2005).

The mass rate transfer of vapour by convection (assuming isothermal conditions and constant dry air pressure $u_{da}$) can be expressed in terms of mixing ratio (mass of vapour per unit mass of dry air) or relative humidity differences between two points in the circuit (in and out) as (Oldecop and Alonso 2004, Dueck 2007).

Figure 7 shows the evolution of vertical strains (expansive deformations are positive) of compacted bentonite subjected to a reduction (from 150 MPa to 4 MPa) in suction under oedometer conditions (vertical net stress of 10 kPa), using both relative humidity controlled chamber (pure diffusion of vapour) and forced flow of humid air on both ends of the sample. As observed, the forced flow speeds up the process of suction change.

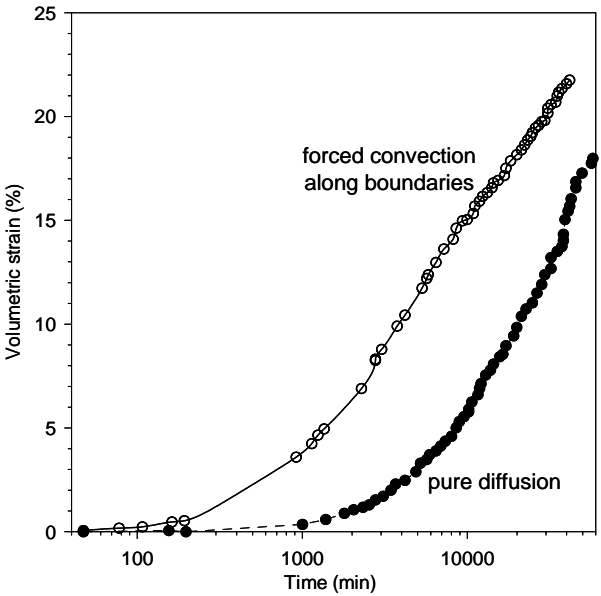

Figure 7. Evolution of volumetric strain on compacted bentonite using humid air flow along the boundaries of the sample (forced convection) or controlling the air relative humidity inside a closed chamber (pure diffusion) (Pintado 2002).

The mass rate transfer of vapour by convection (assuming isothermal conditions and constant dry air pressure $u_{da}$) can be expressed in terms of vapour density or mixing ratio differences between the reference vessel with aqueous solution (superscript $r$) and the soil (superscript $s$) (a: Oldecop and Alonso 2004, b: Jotisankasa et al. 2007)

$$a) \quad M_{dry} \frac{dw}{dt} = q(\rho_v^r - \rho_v^s) = \frac{q M_{mw}}{RT} u_{v0}(h_r^r - h_r^s)$$
$$b) \quad M_{dry} \frac{dw}{dt} = q_{da}(x^r - x^s) = q_{da} x_0 (h_r^r - h_r^s)$$
(4)

where $M_{dry}$ is the soil dry mass, $w$ the gravimetric water content, $q$ the volumetric air flow rate, $\rho_v$ the vapour density in air (water mass per unit volume of air), $q_{da}$ the flow rate of dry air mass, and $x$ the mixing ratio (mass of water vapour per unit mass of dry air; $x_0$ represents the saturated mixing ratio). Assuming vapour an ideal gas, the vapour density can be expressed as $\rho_v = M_{mw} u_{v0} h_r / (RT)$, where $u_{v0}$ is the saturated vapour pressure at absolute temperature $T$, $M_{mw}$ is the molecular mass of water, $R$ is the gas constant, and $h_r$ the relative humidity. Based on the same assumption and that dry air is also an ideal gas, the following expression is obtained $x = M_{mw} u_v / (M_{mda} u_{da}) = 0.622 \, u_v / u_{da}$, in which $M_{mda}$ is the molecular mass of dry air mixture, $u_{da}$ the dry air pressure and $u_v$ the vapour pressure.

One of the difficulties in using the vapour equilibrium technique is associated with maintaining thermal equilibrium between the reference system (vessel with aqueous solution) and the sample. Assuming that the vapour pressure set by the reference saline solution is also present in the sample, the following correction is proposed, in which $h_r$ is the relative humidity and $u_{v0}$ the saturation vapour pressure at temperature $T$

$$h_{r\,sample} = h_{r\,reference} \frac{u_{v0}(T_{reference})}{u_{v0}(T_{sample})} \quad (5)$$

A possible way to minimise this thermal effect is achieved by disconnecting the reference system that regulates the relative humidity, and allow the equalisation of vapour in the remaining circuit and the soil. This way, the mass of water being transferred from or to the soil is drastically reduced (there is no contribution in water transfer between the vessel and the soil). An equivalent testing procedure was used by Oldecop and Alonso (2004) to overcome the long equalisation periods of the conventional vapour equilibrium technique.

Another problem that comes up when using the forced convection system is associated with air pressure differences created along the circuit. This fact makes that the intended relative humidity applied by the reference vessel cannot be assigned to the remaining circuit and the soil. Dueck (2004) studied the influence of air pressure changes in a forced convection circuit of vapour and their consequences on the applied relative humidity. Figure 8 shows the experimental setup and the evolution of differential air pressures between two points of the circuit (before and after the filter stones). The consequences on the evolution of the relative humidity at the same two points of the circuit are shown in Figure 9. An expression to account for the effects of air pressure variations on the relative humidity can be proposed based on the assumption that the mixing ratio $x = 0.622 \, u_v / u_{da}$ (mass of vapour per unit mass of dry air) set by the reference saline solution is also set in the sample under isothermal conditions

$$h_{r\,sample} = h_{r\,reference} \frac{x_{0\,reference}}{x_{0\,sample}}$$

$$h_{r\,sample} = h_{r\,reference} \frac{u_{da\,sample}}{u_{da\,reference}} \quad (6)$$

in which $h_r = x / x_0$ is the relative humidity, $x_0 = 0.622\, u_{v0}/u_{da}$ the saturated mixing ratio ($u_{v0}$ is the saturated vapour pressure), and $u_{da}$ the dry air pressure.

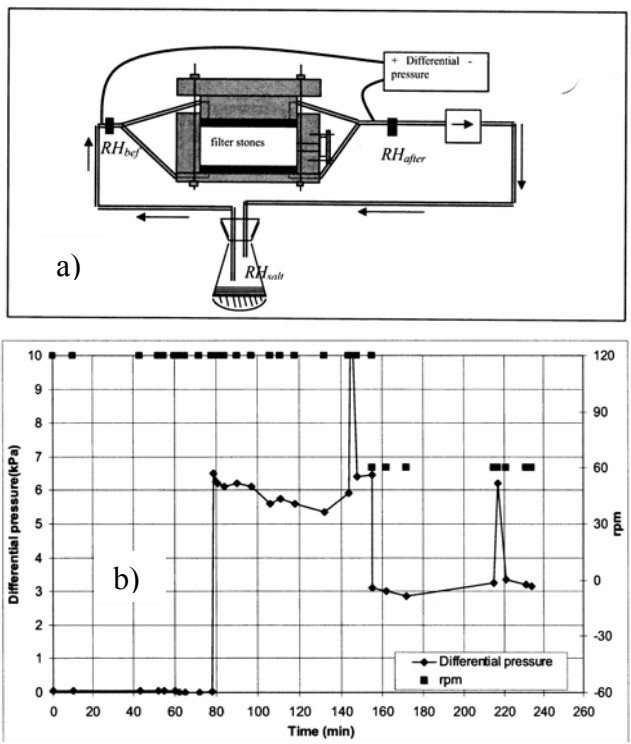

Figure 8. a) Experimental setup to study air pressure and relative humidity changes along a forced convection circuit. b) Time evolution of differential air pressures (at 80 min the air is forced through the filter stones) (Dueck 2004).

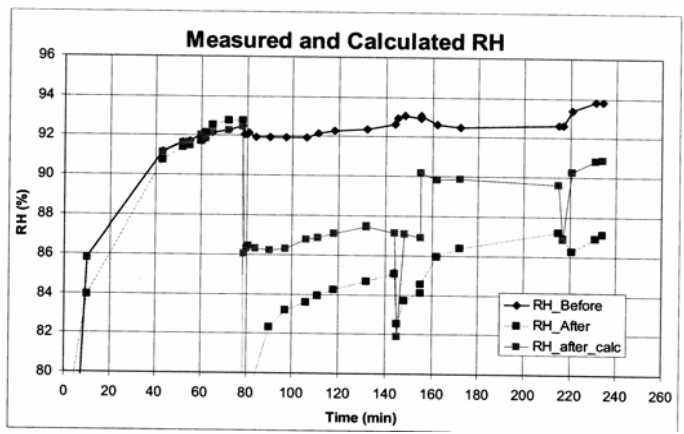

Figure 9. Time evolution of measured and calculated relative humidity at two points of the forced convection circuit (at 80 min the air is forced through the filter stones) (Dueck 2004).

## 3 TECHNIQUES OF MEASURING SUCTION

### 3.1 High capacity tensiometers

In terms of suction measurement, significant progress has been made with the development of the high-capacity tensiometer (HCT) by Ridley & Burland (1993). The HCT is similar in conception to a standard tensiometer and comprises a water reservoir, a high air-entry interface and a pressure gauge. Figure 10 shows the second prototype developed by Ridley & Burland (1995) which will be referred to as 'IC tensiometer'. This tensiometer includes an integral strain-gauged diaphragm in contrast to the first prototype (Ridley & Burland 1993) obtained by fitting a porous ceramic disk to a commercial pressure transducer. Key elements of the IC tensiometer were the very thin water reservoir (less than 4 mm$^3$) and the use of a sufficiently thick high air-entry value ceramic disk (Ridley 1993). Provided adequate de-airing processes and pressurisation were carried out, the IC tensiometer could move the maximum sustainable suction up to 1800 kPa, a value significantly higher than 70-80 kPa typical of standard tensiometers.

The IC tensiometer was particularly welcome by the geotechnical community since suction was difficult to measure accurately in the range 0-1500 kPa using other techniques such as the psychrometer. As shown in

Table 1, the concept of the IC tensiometer had significant success and many similar devices have been developed since that time with some specific technological improvements, as discussed in Tarantino (2004), Mahler and Diene (2007), and Marinho et al. (2008). Design and use of high capacity tensiometers have been satisfactorily documented. This measurement technique now appears to be reasonably affordable to develop and to use in the laboratory.

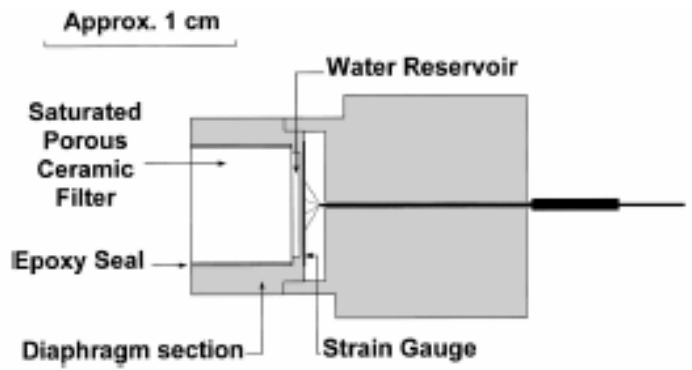

Figure 10. The Imperial College suction probe (Ridley and Burland 1995).

Table 1. High-capacity tensiometers developed by various authors including the pressure transducers used.

| Authors | Pressure transducer |
| --- | --- |
| Ridley & Burland (1993) | Entran EPX (3.5 MPa) |
| König et al. (1994) | Druck PDCR 81 (1.5 MPa) |
| Ridley & Burland (1995) | Home-made (4 MPa) |
| Guan & Fredlund (1997) | Brand not given (1.5 MPa) |
| Meilani et al. (2002) | Druck PDCR 81 (1.5 MPa) |
| Tarantino & Mongiovi (2002) | Home-made (4 MPa) |
| Take and Bolton (2002, 2003) | Druck PDCR 81 (1.5 MPa) and Entran EPB (0.7 MPa) |
| Toker et al. (2004) | Data Instr. Inc. AB-HP 200 |
| Mahler et al. (2002) | Ashcroft K8 |
| Chiu et al. (2005) | Druck PDCR 81 (1.5 MPa) |
| Lourenco et al. (2006, 2007) | Ceramic transducer by Wykeham Farrance (0.8 MPa) |
| Oliveira and Marinho (2007) | Entran EPX (3.5 MPa) |
| He et al. (2006) | Entran EPX (3.5 MPa) |
| Mahler & Diene (2007) | Entran EPX (1.5 MPa) Entran EPXO (0.5 MPa) Ashcroft (0.5-1.5 MPa) |
| Cui et al. (2008) | Home-made |

### 3.1.1 Water under tension and cavitation

Cavitation of water typically occurring at negative gauge pressures close to -70/-80 kPa has long been explained by the inability of water to sustain tensile stresses. This supposition is incorrect as water can indeed sustain high tensile stresses as earlier recognised by Berthelot (1850) and confirmed by several experiments carried out by using metal and glass Berthelot-type systems (see Marinho and Chandler 1995). Using a Berthelot-type device, relatively long measurements could be carried out by Henderson & Speedy (1980) who reported a tension of 10 MPa sustained for over a week. Zheng et al. (1991) were able to measure a tensile stress of 140 MPa in a single crystal of water, a value believed to be very close to the maximal tension that water can sustain.

The state of water under tension is thermodynamically metastable (De Benedetti 1996) in the sense that a gas phase will rapidly separate in the liquid if tiny amounts of gas (cavitation nuclei) are pre-existent in the liquid. Marinho and Chandler (1995) reviewed the sources of impurities in the water which include i) solid particles that contain gas micro-bubbles trapped in crevices, ii) gas trapped in tiny crevices in the walls of the water container, iii) air bubbles stabilized by ionic phenomena and iv) bubbles covered by surface active substances. Note that in high range tensiometers, case i) also applies to the pores of the ceramic porous stone.

Since it is virtually impossible to completely remove air from the water reservoir and the porous ceramic filter, heterogeneous cavitation will inevitably occur in the HCTs. The main challenge in tensiometer measurement is then to delay cavitation by minimising the number of potential cavitation nuclei present in the tensiometer. This has essentially been achieved by adopting special design features and by implementing specific procedures for saturating the porous ceramic disk (initial saturation and subsequent re-saturation).

### 3.1.2 Design

The very small water reservoir designed by Ridley and Burland (1993, 1995) was assumed to decrease the number of cavitation germs in free water and hence the probability of cavitation occurrence. In this regard, Ridley and Burland (1999) mentioned that the change in design from the 1993 to the 1995 IC tensiometer was aimed at reducing as far as possible the size of the water reservoir. According to their experience, this reduction (with a water reservoir thickness close to 0.1 mm as shown in Figure 10) appeared to allow suction measurements with no cavitation along a longer period of time, with less random breakdowns of the measurements (Guan and Fredlund 1997 give a water reservoir thickness between 0.1 and 0.5 mm). Reducing the thickness of the water reservoir is believed to be an important feature necessary to develop high capacity tensiometers that has been followed in all the prototypes described in

Table 1. In general, the reservoir volume is of the order of 5-10 mm$^3$ with thickness as low as 0.1 mm.

Two types of design have been presented in the literature, integral strain-gauged tensiometers (Ridley & Burland 1995, Tarantino & Mongiovì 2002, Cui et al. 2008) and tensiometers obtained by fitting a high AEV ceramic disk to a commercial transducer. The latter can be further divided in three classes, depending on whether the water reservoir was sealed by means of O-Ring (Ridley & Burland 1993, Guan & Fredlund 1997, He et al. 2006), araldite (Meilani et al. 2002, Take & Bolton 2003, Lourenço et al. 2006) or copper gasket (Toker et al. 2004).

In general, the best performance in terms of maximum sustainable tension and measurement duration appear to be achieved by the integral strain-gauged diaphragms. On the other end, concerns arise about the use of O-rings to seal the water reservoir. The change in design from the 1993 to the 1995 IC tensiometer was also aimed at eliminating O-rings and elastomers which are sources of nucleation sites (Take 2003). Toker et al. (2004) also found that cavitation occurred at very low tensions when sealing the water reservoir using rubber O-rings and that significant improvement could be obtained by replacing the O-ring with araldite or copper gasket. The tensiometer presented by Guan & Fredlund (1997) which included an O-ring to seal the water reservoir also exhibited relative poor performance. Despite the high pre-pressurisation pressure (12 MPa), the maximum sustained tension (1.25 MPa) was significantly lower than the nominal

AEV of the ceramic disk (1.5 MPa). As shown in the next section, this is not the case of integral strain-gauged diaphragms and araldite-assembled tensiometers where maximum sustained tension can significantly exceed the nominal AEV of the ceramic disk.

### 3.1.3 Initial saturation

Ridley and Burland (1999) emphasized the importance of careful initial saturation of the porous stone by de-aired water under vacuum, prior to pressurisation. They observed that a subsequent pressurisation at 4 MPa for at least 24 h could provide satisfactory suction measurements. Adopting these precautions, they concluded that the maximum sustainable suction was only depending on the air entry value of the ceramic filter, most often equal to 1500 kPa in existing devices. This observation is nicely illustrated by the results presented in Figure 11 (Ridley and Burland 1999) that shows the maximum suction obtained with various ceramic porous stones with air entry values (AEV) of 100, 500 and 1500 kPa respectively.

It is interesting to note that the combination of an initial saturation under vacuum and a pre-pressurisation pressure about 2.7 times the AEV of the ceramic disk could produce maximum sustained tensions significantly higher than the nominal AEVs of the porous ceramic disks (164/100 kPa, 740/500 kPa and 1800/1500 kPa respectively).

The importance of the initial saturation under vacuum has also been discussed by Take & Bolton (2003). Three procedures for initial saturation of the ceramic disk were investigated i) saturation at atmospheric pressure; ii) evacuation in presence of water followed by saturation under vacuum; iii) evacuation in absence of water followed by saturation under vacuum. In case i), the tensiometer could not sustain any tension even after four pre-pressurisation cycle of 1000 kPa. In case ii), once subjecting the tensiometer to a pre-pressurisation cycle of 1000 kPa, a maximum sustainable tension of 460 kPa could be attained (greater that the nominal 300 kPa AEV of the ceramic disk). In this case, vacuum in presence of water was somehow limited by the vapour pressure of water (2.3 kPa at 20°C). Finally, in case iii), an absolute pressure of 0.05 kPa could be attained when applying vacuum in absence of water and the maximum sustainable tension, once subjecting the tensiometer to a pre-pressurisation cycle of 1000 kPa, could be increased to 530 kPa.

A procedure similar to case iii) was devised by Tarantino & Mongiovì (2002) with the exception of the porous ceramic initially dried using silica gel instead of oven-drying at 60°C.

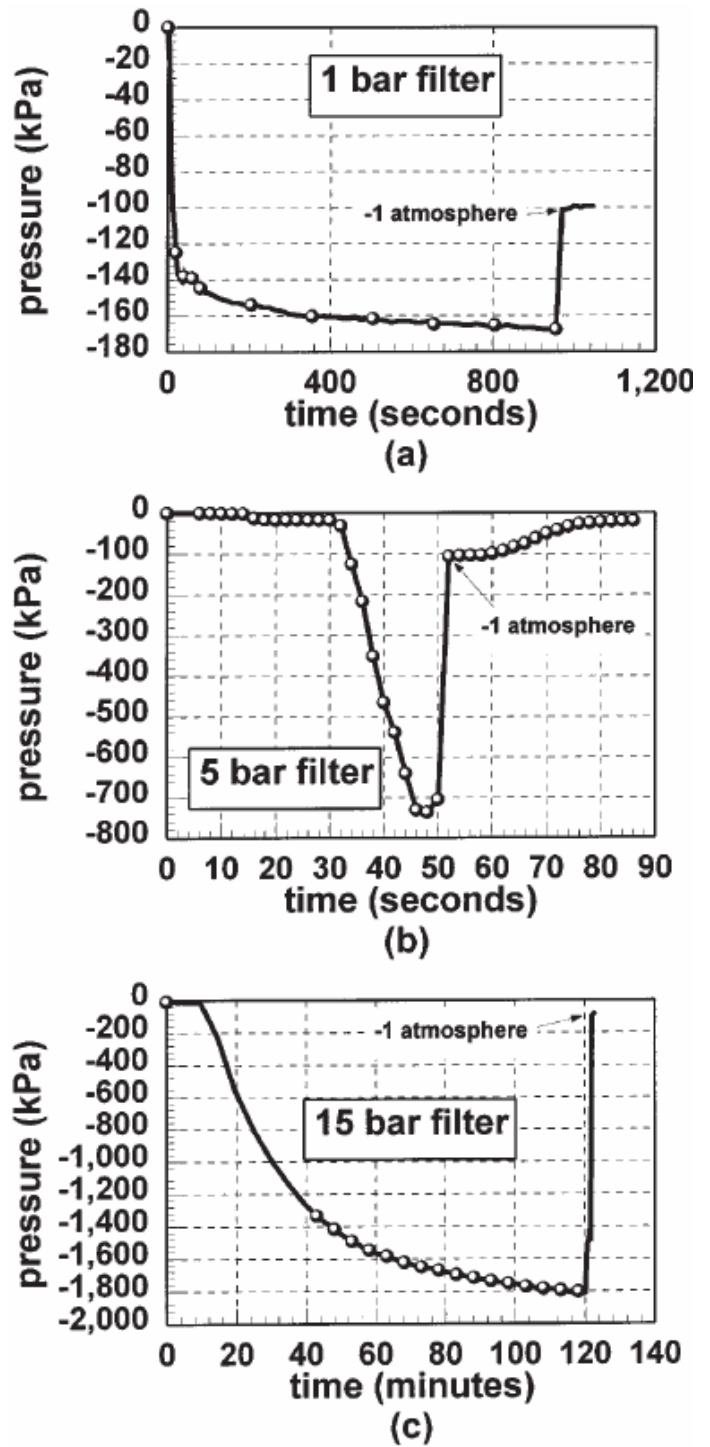

Figure 11. Maximum suction response obtained with various ceramic porous stones (Ridley and Burland 1999).

### 3.1.4 Pre-pressurisation

An issue that has long been debated is the procedure to be used to re-saturate the porous ceramic disk. Guan and Fredlund (1998) observed that the cavitation tension was essentially depending on the number of pre-pressurisation cycles and to a less extent on the pre-pressurisation pressure and duration. In particular, they found that 6 pressures cycles from -0.1 to 12 MPa produced the maximum sustainable tension. Cycles including the application of a positive pressure followed by a negative gauge pressure of about -0.1 MPa were also used by Take & Bolton (2003).

On the other hand, Ridley & Burland (1995) observed that, provided initial saturation was carried out under vacuum, pre-pressurisation at a constant pressure of 4 MPa (2.7 times the nominal AEV) for a period of 24 h was sufficient to measure water tensions higher than the ceramic disk nominal AEV. The application of a constant pre-pressurisation pressure over a period of time was also adopted by Tarantino & Mongiovì (2002), Meilani et al. (2002), Chiu *et al.* (2005), Lourenço et al. (2006), He *et al.* (2006) and Cui *et al.* (2008).

There is no experimental evidence showing that that one procedure is preferable to the other. On the other hand, little attention has been given so far to the pre-pressurisation pressure in relation to the AEV of the ceramic disk. Table 2 shows the pre-pressurisation pressure adopted by different authors together with the AEV of the ceramic disk and the maximum sustained tension. It can be observed that pre-pressurisation pressures greater than 2.7 times the nominal AEV of the ceramic disk can produce maximum tensions greater than the ceramic AEV (Ridley & Burland 1995, Tarantino & Mongiovì 2002, Take & Bolton 2003, He *et al.* 2006).

Table 2. Effect of the pre-pressurisation pressure on the maximum sustained tension (in bold sustained tension greater than the ceramic disk nominal AEV)

| Authors | Ceramic AEV (MPa) | Max positive pressure (MPa) | Max water tension (MPa) |
| --- | --- | --- | --- |
| Ridley & Burland (1993) | 1.5 | 6 | 1.37 |
| Ridley & Burland (1995) | 0.1 | 4 | **0.164** |
|  | 0.5 | 4 | **0.74** |
|  | 1.5 | 4 | **1.8** |
| Guan and Fredlund (1997) | 1.5 | 12 | 1.25 |
| Meilani *et al.* (2002) | 0.5 | 0.8 | 0.495 |
| Tarantino & Mongiovi (2002) | 1.5 MPa | 4 MPa | **2.06** |
| Take & Bolton (2003) | 0.3 | 1 | **0.53** |
| Mahler *et al.* (2002) |  |  |  |
| Chiu *et al.* (2005) | 0.5 | 0.7 | 0.47 |
| Lourenco *et al.* (2006) | 1.5 | 1 | 1.23 |
| He *et al.* (2006) | 0.5 | 2 | **0.55** |
| Mahler and Diene (2007) | 0.5 | 0.6 | **0.8** |
|  | 1.5 | 0.6 | 1.4 |

The only exception is given by the tensiometers developed by Ridley & Burland (1993) and Guan & Fredlund (1997) which, however, were sealed using O-rings. On the other hand, pre-pressurisation pressures in the range 0.67-1.6 AEV appear to produce relative poor performance (Meilani et al. 2002, Chiu et al. 2005, Lourenço *et al.* 2006), in the sense that the maximum sustained tension was lower than the ceramic disk AEV. The only exception appears to be given by the tensiometers by Mahler & Diene (2007) which, however, showed unusual response upon cavitation as water pressure appear to return to zero instead of -100 kPa gauge pressure as in all other tensiometers presented in the literature.

Another experimental procedure that can be adopted to improve both maximum sustainable tension and measurement duration consists in subjecting the tensiometers to repeated cycles of cavitation (induced by placing the probe in contact with a dry sample for instance) and subsequent pressurisation. Experimental evidence of the beneficial effect of this procedure is provided by Tarantino & Mongiovì (2001) using IC tensiometer, Tarantino & Mongiovì (2002) using Trento tensiometer, Toker (2002) using MIT 6.1 tensiometer, and Take & Bolton (2003) using the tensiometer developed at the University of Cambridge. The application of repeated cycles of cavitation and pre-pressurisation appeared to not improve the response of the tensiometer presented by Chiu et al. (2005) and Lourenço et al. (2006). However, these authors applied relatively low pre-pressurisation pressures (see Table 2) which may explain the non-beneficial effect of this procedure.

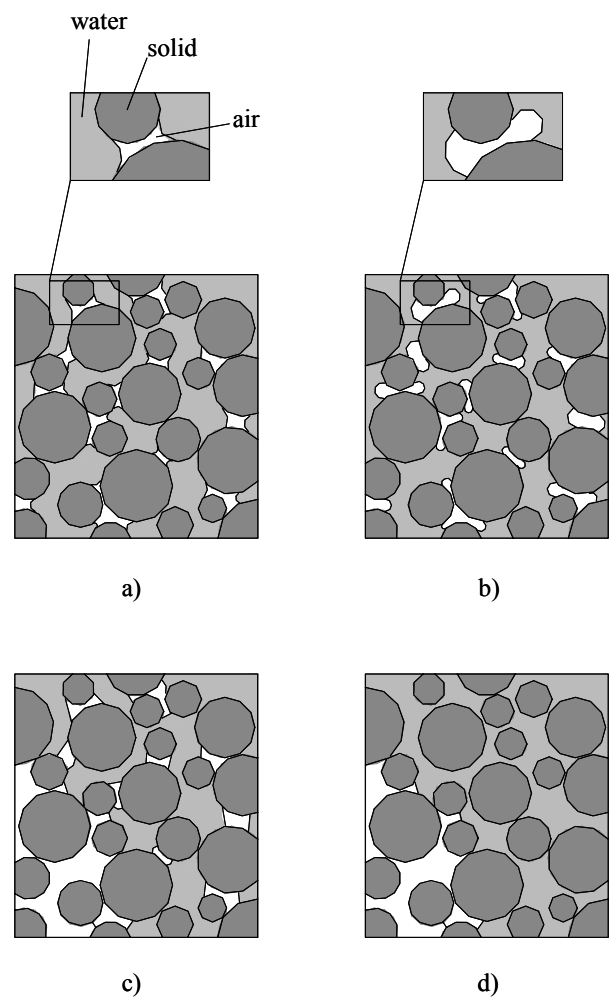

Figure 12. Possible cavitation mechanism inside the tensiometer: a) pre-pressurisation; b) measurement; c) cavitation; d) air diffusion (Tarantino and Mongiovi (2001)

Tarantino & Mongiovì (2001) assumed that repeated cycles of cavitation and re-saturation can reduce number and size of cavitation nuclei in the po-rous ceramic disk. Air in small nuclei would be driven together by cavitation in larger cavities that are subsequently more easily forced into solution by pressurisation. This phenomenon is illustrated in Figure 12 and would suggest that cavitation occurs inside the porous stone rather that in the water reservoir. Experimental evidence supporting this assumption is provided by Tarantino & Mongiovì (2001) and Tarantino (2004). Evidence is also given by Guan and Fredlund (1997) who observed that the inner face of the ceramic disk became relatively soft after repeated cavitations and it was possible to peel the surface with a slight fingernail scratch. This degradation is likely to be related to the occurrence of cavitation localised in this area.

### 3.1.5 *Evaluation of tensiometer performance*

The quality of the suction measurements provided by the various tensiometers developed so far has been investigated by various means by the different authors. Most often, the tensiometer calibration has been conducted by extrapolating calibration established with positive water pressures to negative pressures. In a first attempt, Ridley and Burland (1993) performed suction measurements on a saturated clay sample put under a given isotropic effective stress situation and subsequently unloaded in undrained conditions, considering that a suction equal to the effective stress would develop (hence implicitly assuming sample isotropy and "perfect sampling", see Doran et al. 2000). Guan and Fredlund (1997) measured the suction of samples put at controlled suctions by using the axis translation method. They also compared the tensiometer measured suctions with filter paper measurements, as done also by Marinho and Chandler (1994) in an attempt to investigate possible osmotic effects. Tarantino and Mongiovi (2001) stated that the comparison between direct and indirect methods was probably not the best approach, and they successfully compared the measurements given by two IC tensiometers put in contact with the same sample, concluding on the satisfactory quality of the measurement. They also observed an excellent agreement between the measurements given by a Trento tensiometer and a IC tensiometer placed in contact with a dry kaolinite sample, before cavitation (Tarantino and Mongiovi 2002). Such a good agreement between two different systems is certainly a good indicator of the quality of the measurement. Note that in this experiment, the IC tensiometer was able to reach a suction value as high as 2900 kPa before cavitating.

### 3.1.6 *Time to reach equilibrium*

All authors agree that the contact between the soil and the probe needs particular attention, a good contact being ensured by placing a soil paste between the probe and the sample. However, the water content of the paste may affect the time necessary to reach equilibrium. Oliveira and Marinho (2008) used soil pastes at various water contents and recommended to choose the water content between the plastic and liquid limits, with higher water contents resulting in longer equilibration rates. Boso et al. (2004) also showed that preparing the paste at the liquid limit may significantly increase the equilibration time. They suggested that the water content of the paste should be kept as low as possible. However contact may not establish if the paste water content is excessively low and optimal water content should therefore be chosen by trial and error.

Oliveira and Marinho (2008) also showed that equilibration time also depends on the permeability of the soil. Figure 13 shows suction measurements on samples of low plasticity soil ($I_P = 13$) compacted on the dry side, wet side and at Proctor optimum. The results show that similar levels of suction (between 400 and 500 kPa) are attained after different periods of time due to changes in the microstructure.

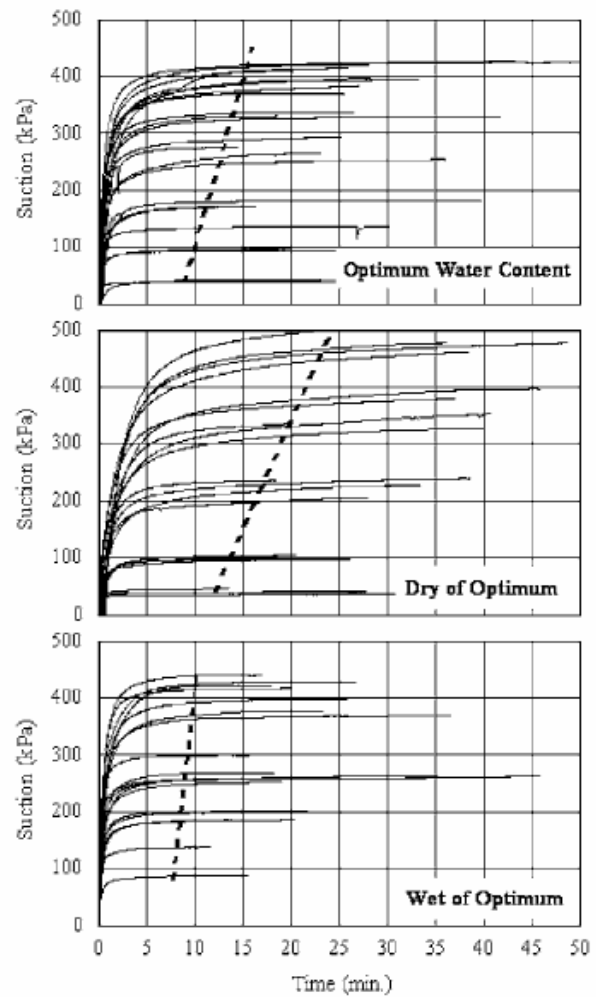

Figure 13. Effect of compacted sample microstructure on suction measurements (Oliveira and Marinho (2008).

The changes in permeability of compacted samples with the compaction state has been known for long time since the work of Lambe (1958) who observed a decrease of two orders of magnitude when passing from the dry side of optimum to optimum water content, followed by a constant value along the saturation line. This is compatible with the trend observed in Figure 13 in which similar response are observed at optimum and on the wet side of optimum, as compared to longer equilibration times on the dry side. This is related to the aggregate microstructure observed on the dry side (Ahmed et al. 1974, Delage et al. 1996, Romero et al. 1999), as compared to the matrix microstructure on the wet side. Obviously, the measurement of suction is corresponding to a very tiny water movement that is sufficient to extract some water from the porous stone. This transfer rate is dependent on the microstructure, with slower rates in the aggregate macrostructure, in which inter-aggregates pores are known to be dry and in which water is moving through the inter-aggregates contacts and, probably, inside the inter-aggregates smaller pores.

### 3.1.7 *Use in geotechnical testing*

Tensiometers have been extensively used in mechanical testing including null tests (Tarantino *et al.* 2000), oedometer tests (Dineen & Burlan 1995, Dineen et al.,1999, Tarantino & Mongiovì 2000, Delage *et al*. 2007, Tarantino & De Col 2008), direct shear tests (Caruso & Tarantino 2004, Tarantino & Tombolato 2005), and triaxial tests (Cunningham *et al.* 2003, Oliveira & Marinho 2003).

The use of the tensiometer makes it possible to investigate unsaturated soil behaviour under more realistic atmospheric conditions. Tests have been carried out under suction-controlled conditions by coupling the tensiometer with either the osmotic technique (Dineen & Burland 1995, Dineen et al. 1999, Tarantino & Mongiovì 2000) or air circulation (Cunningham et al. 2003). Tests have also been carried out at constant water content with suction changes monitored by the tensiometer (Tarantino & Tombolato 2005, Delage *et al*. 2007, Tarantino & De Col 2008). It is interesting to observe that suction equalisation in constant water tests is generally fast (1-2 h) which significantly reduces the overall test duration. This point is of importance especially if a comparison is made with the long-lasting tests based on the axis-translation technique.

The advantage of using the tensiometer is that quasi-saturate states (occluded air-phase) and the transition from unsaturated to saturated states can be successfully examined in contrast to the axis-translation technique which is problematic to use at very high degrees of saturation. For example, Tarantino & Mongiovì (2000) could perform constant suction one-dimensional compression on a sample having an initial degree of saturation equal to 0.95.

As seen in Figure 14, Delage *et al.* (2007) could monitor the changes in suction or water pressure occurring during a step loading oedometer compression test carried out on a saturated intact Boom clay sample (a stiff clay from Belgium) using dry porous stones. The initial suction state of the saturated sample resulted from the stress release due to sample extraction (block sampling). In the range where measured pressures remain negative (vertical stress smaller than 800 kPa), the figure shows that each loading step results in a peak in the response of the tensiometer, with apparently positive pressures monitored just before going back to a suction state. These peaks, not always equal to the stress increment applied, are interpreted as a local consolidation process of a thin soil layer in contact with the bottom of the cell where the tensiometer was placed. This instantaneous positive response is apparently very quickly compensated by suction subsequent homogenisation within the soil mass. Note that the transition between negative and positive pressures is well captured once a load of 800 kPa is reached, with subsequent stabilisation of the pressure measurements at zero. During unloading, a suction state seems reached again when the load is smaller than 400 kPa.

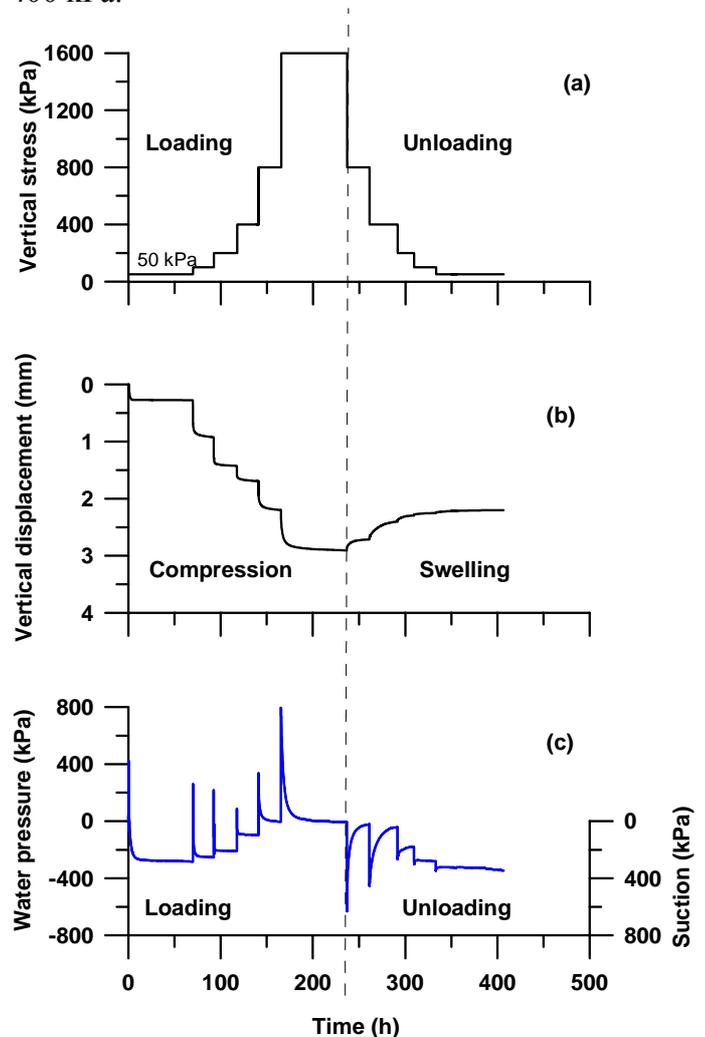

Figure 14. Monitoring suction changes during oedometer step loading compression (After Delage et al. 2007)

As seen in Figure 15, Tarantino & De Col (2008) could investigate suction changes occurring during the compaction process in clay samples at various water contents. The figure clearly shows the simul-taneous decrease in suction and increase in degree of saturation that occured during compaction. Also apparent are the hysteresis obtained during stress cycles and the significant suction increase due to vertical stress release. The profile of the post-compaction suctions at given water content, once stress is released, shows some increase in suction with increasing the degree of saturation. This trend is different from some observations by Li (1995), Gens et al. (1995) and Romero et al. (1999) who measured constant values of suction on samples of equal water content compacted at various densities, at least on the dry side of the compaction curve.

When implementing the HCTs in mechanical testing, an important aspect is that water tension has to be sustained for a time long enough to carry out the test. Tarantino & Mongiovì (2000) and Cunningham et al. (2003) showed that water tensions of the order of 800 kPa could be sustained for more than two weeks without cavitation occurring in the tensiometer. Interestingly, water tensions were simultaneously measured using two tensiometers which showed excellent agreement with each other.

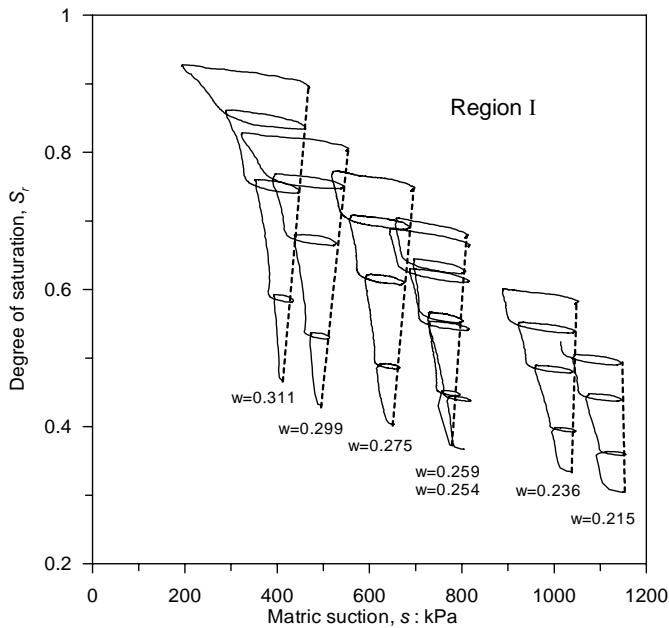

Figure 15. Degree of saturation-suction paths at different compaction water contents. Dotted lines join "post-compaction" suctions (Tarantino and De Col 2008).

The results and considerations presented in this section devoted to the direct measurement of suction by high capacity tensiometers clearly show the quality of the measurements obtained with this device provided adequate preliminary preparation procedures are carried out. Further use of HCTs in laboratory testing of unsaturated soils will definitely complete in a sound fashion our understanding of the behaviour of unsaturated soils.

### 3.1.8 *In-situ suction measurements*

The measurement of in-situ suction profiles and of suction changes with respect to time and climatic changes is certainly an important aspect in which progress is needed in the mechanics of unsaturated soils, with obvious applications in many fields in which soil atmosphere exchanges are playing a key role. The behaviour and stability of geotechnical structures like embankments or earth-dams, cover liners of surface waste disposals and the investigation of slope stability problems are some typical examples.

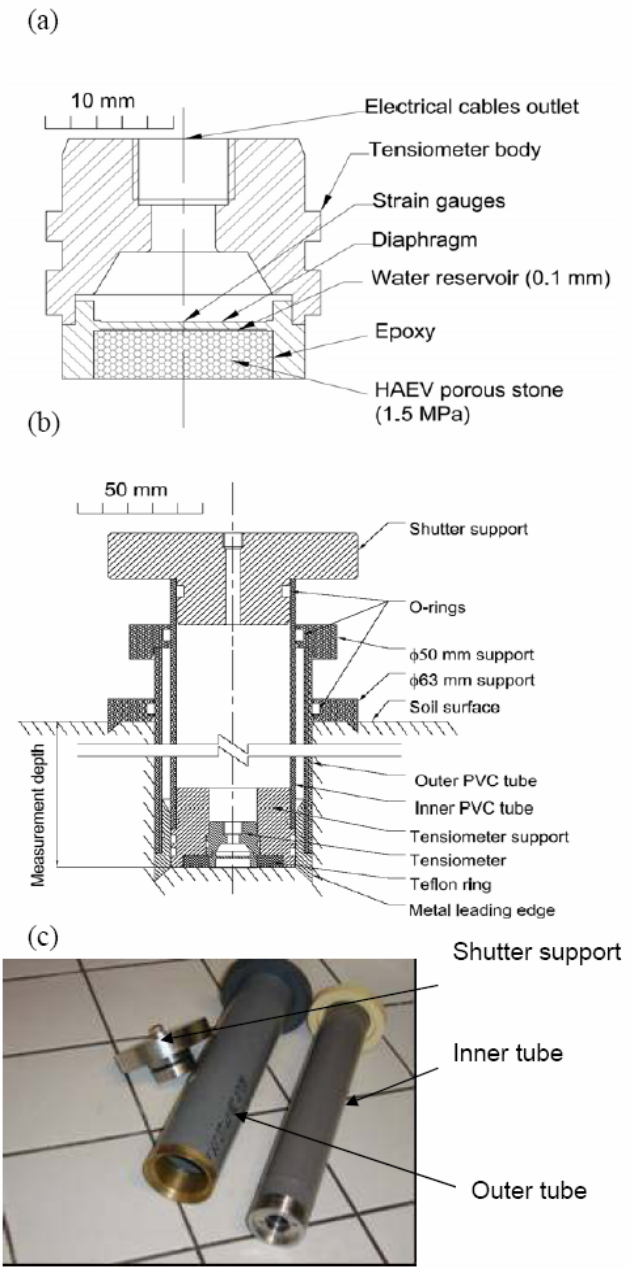

Figure 16. In-situ measurement of soil suction at shallow depth (Cui et al. 2008)

In spite of some attempts (Ridley et al. 1996), it seems that experimental in-situ suction profiles determined by using high capacity tensiometers are scarce. In this regard, a recent paper by Cui et al. (2008) shows a device allowing to measure suction changes in a low range (20-160 kPa) at small depths

(25 cm and 45 cm) along a period of three weeks. The system (Figure 16) also allows simple replacement of the tensiometer to carry out, when necessary, a new resaturation of the probe in the laboratory after the occurrence of cavitation due to progressive air diffusion in the porous stone. The authors are aware that the period of three weeks measurements should probably be reduced when measuring higher in-situ suctions.

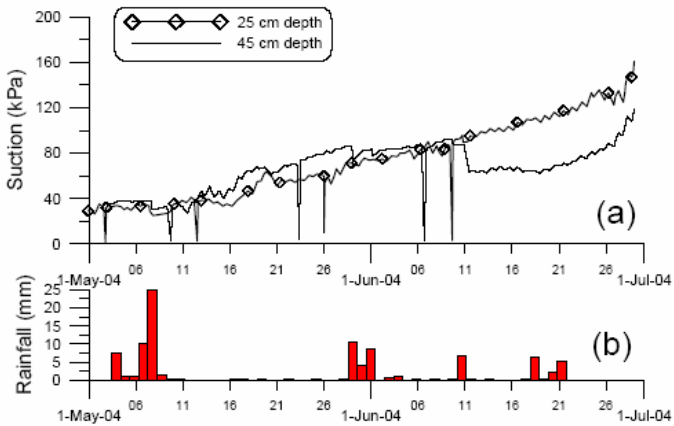

Figure 17. Suction changes and rainfall data (Cui et al. 2008).

The data of Figure 17 show that rainfall events generally cause slight changes in suction, with a more pronounced reaction at 45 cm on 11 June.

Due to progressive air diffusion in the ceramic disk, it is not sure that direct measurements of in-situ suction by using HCT be the more reliable technique to be used for long term monitoring, as compared to indirect techniques (see for instance Whalley et al. 2007).

3.2 *High-range psychrometers*

In recent years, a great effort has been dedicated to extend the working range of psychrometers. Improved versions of widely used thermocouple psychrometers (Peltier-type: Spanner 1951; wet-loop: Richards and Ogata 1958; double-junction: Meeuwig 1972, Campbell 1979; screen-caged: Brown and Johnston 1976, Brown and Collins 1980) displayed a range between 0.3 MPa and 8 MPa, although above 4 MPa the repeatability of the outputs was not very good (Ridley and Wray 1996). Two alternatives, with different working principles, have been developed in the 90s and 2000s that allow extending this range: a) transistor psychrometers (Soil Mechanics Instrumentation SMI type: Dimos 1991, Woodburn et al. 1993, Truong and Holden 1995) with an upper limit of 15 MPa, and b) chilled-mirror dew-point psychrometers (Gee et al. 1992, Loiseau 2001, Brye 2003, Leong et al. 2003, Tang and Cui 2005, Thakur and Singh 2005, Agus and Schanz 2005) with an upper limit of 60 MPa and involving a reduced time of reading. Woodburn and Lucas (1995) and Mata et al. (2002) extended the range of transistor psychrometers to 70 MPa, disconnecting the probes from the standard logger and reading the outputs using a millivoltmeter.

The transistor psychrometer probe consists of two bulbs, which act as 'wet' and 'dry' thermometers that are placed inside a sealed and thermally insulated chamber in equilibrium with soil sample. A drop of distilled water with specified dimensions is used in the 'wet' thermometer. The psychrometer measures indirectly the relative humidity by the difference in temperature between the 'dry' and the 'wet' bulbs (evaporation from the 'wet' bulb lowers its temperature). The standard equilibration period is one hour. The extended range calibration for standard conditions (standard drop size and equilibration period) is bi-linear, as shown by Cardoso et al. (2007). On the other hand, the chilled-mirror dew-point psychrometer measures the temperature at which condensation first appears (dew-point temperature). A soil sample in equilibrium with the surrounding air is placed in a housing chamber containing a mirror and a photoelectric detector of condensation on the mirror. The temperature of the mirror is precisely controlled by a thermoelectric (Peltier) cooler. The relative humidity is computed from the difference between the dew-point temperature and the temperature of the soil sample, which is measured with an infrared thermometer. The measuring time is around 5 minutes (WP4 psychrometer, Decagon Devices, Inc.) Table 3 presents the comparison of both equipment (SMI and WP4) concerning suction range, output, accuracy, measurement time and calibration.

Table 3: Specifications of two high-range psychrometers (Cardoso et al. 2007).

| Equipment | SMI Psychrometer | Chilled-mirror dew-point WP4 |
|---|---|---|
| Suction range | 1 to 70MPa (*) | 1 to 60MPa (max. 300MPa) |
| Output reading | Voltage, suction (logger) | Suction and temperature |
| Accuracy | <±0.05 pF ±0.01 pF (repeatability) | ±0.1MPa from 1 to 10MPa and ±1% from 10 to 60 MPa |
| Measuring time | Usually 1 hour | 3 to 10 minutes |
| Calibration | Multiple point calibration. Bi-linear | Single point calibration |
| Sample geometry | Ø=15mm, h=12mm | Sample cup: Ø=37mm, h=7mm |

Figure 18 shows multi-stage drying results of a clayey silt obtained with SMI psychrometers within this extended range (Boso et al. 2003). The figure also includes results obtained with high-range tensiometer readings. The 'dynamic' determination in the figure was monitored continuously by placing the sample along with the tensiometer on a balance.

The 'static' determination was achieved by constant water content measurements. To compare matrix suction results, a constant osmotic suction of 0.3 MPa was subtracted from total suctions measured by the psychrometer. A relatively good overlapping in the range from 1 MPa to nearly 3 MPa and between the different techniques is observed in the figure.

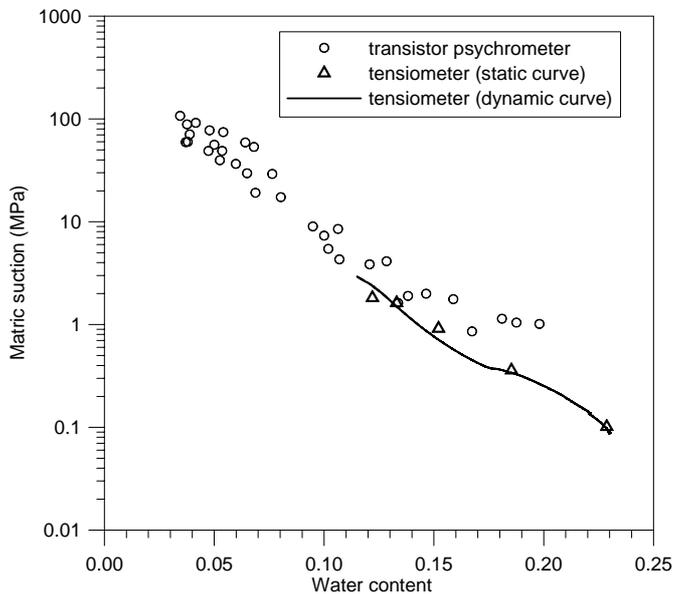

Figure 18. Comparison between SMI psychrometer data (total suction minus osmotic component) and high-range tensiometer readings. Drying paths on a clayey silt (Boso et al. 2003).

Cardoso et al. (2007) studied the performance of SMI and WP4 psychrometers by evaluating the drying branch of the retention curve of a compacted destructured argillite.

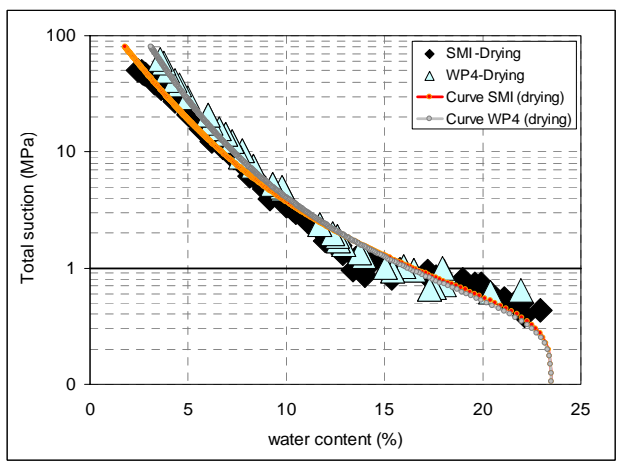

Figure 19. Comparison between SMI and WP4 psychrometer data. Drying paths on a compacted destructured argillite (Cardoso et al. 2007).

As observed in Figure 19, the retention curves display a quite good agreement in the low total suction range from 1 to 7 MPa. However, in the high-suction range (7 to 70 MPa) differences between the readings of both psychrometers were observed – systematically larger values were detected with WP4 psychrometer–, which increased with total suction of the soil.

Cardoso et al. (2007) put forward a possible explanation to account for these discrepancies between SMI and WP4 readings. These authors suggested that the hydraulic paths undergone by the soil during the measurement period inside each equipment chamber were quite different. As observed in Figure 20, the sample in the SMI chamber experiences some wetting due to the relatively fast evaporation of the drop of the wet thermometer, which increases the relative humidity of the chamber to $HR_1 > HR_0$ as shown schematically in the figure. The sample at a lower relative humidity $HR_{soil}$ undergoes some wetting before reaching the equalisation state at $HR_{eq\ SMI}$, which is the state finally measured by the SMI psychrometer. During the determination of a main drying curve, SMI readings will follow a scanning wetting path, which will end below the main drying curve. On the contrary, the soil inside the WP4 chamber will undergo some drying before reaching $HR_{eq\ WP4}$, and it will follow the same intended main drying path during the measuring period. As a consequence, the total suctions measured and the final water contents are slightly different.

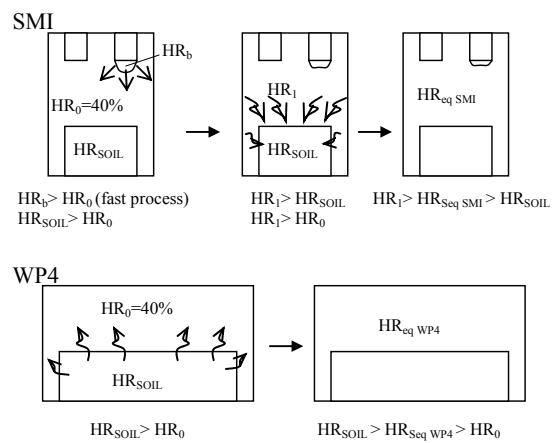

Figure 20. Equalisation process in the measurement chamber of SMI and WP4 psychrometers. $HR_b$: relative humidity near the wet bulb; $HR_{SOIL}$: relative humidity of the soil pores; $HR_0$: initial relative humidity of the soil chamber; $HR_1$: intermediate relative humidity; $HR_{eq}$: final equilibrated relative humidity. The scheme is for $HR_{SOIL} > HR_0$ (Cardoso et al. 2007).

## 4 CONCLUSION

Some recent developments concerning the three techniques used for controlling suction in unsaturated soils (axis-translation, osmotic and vapour control techniques) and concerning two techniques of measuring suction (high capacity tensiometers and high range psychrometers) have been commented and discussed. The advantages, drawbacks and complementarities of these techniques have been discussed and some recommendations aimed at facilitating their use have been given, based on the experience gained by the authors, their co-workers and data available in the literature. As a general conclusion, it can be stated that the recent significant progresses made in the field of controlling and measuring suction provided further insight into the behaviour of unsaturated soils. The potentialities of these techniques are high and they should keep helping the experimental investigations necessary to better understand the hidden remaining aspects of the hydromechanical behaviour of unsaturated soils.

## 5 ACKNOWLEDGEMENTS

The authors acknowledge the fruitful collaboration and discussions with the many colleagues involved in the works conducted: C. Airò Farulla, M. Boso, R. Cardoso, A. Caruso, Y.J. Cui, E. De Col, V. De Gennaro, E. De Laure, A. Di Mariano, A. Dueck, A. Ferrari, Ch. Hoffmann, M. Howat, T.T. Le, A. Lima, A. Lloret, C. Loiseau, A.T. Mantho, D. Marcial, F. Marinho, L. Mongiovi, L. Oldecop, X. Pintado, G. Priol, G.P.R. Suraj de Silva, A. Take, A.M. Tang, A. Thielen, S. Tombolato, T. Vicol, M. Yahia-Aissa.

The authors also wish to acknowledge the support of the European Commission via the "Marie Curie" Research Training Network contract number MRTN-CT-2004-506861.